\documentclass[longauth]{aa}

\usepackage{graphicx}
\usepackage{txfonts}
\usepackage{amsmath}
\usepackage[]{hyperref}

\hypersetup{
    colorlinks = true,
    linkcolor = {blue}, 
    citecolor = {blue},
    urlcolor = {blue}
}

\usepackage{natbib}
\usepackage{color}
\usepackage{float}
\usepackage{xspace}

\newcommand{\spitzer}{\textsl{Spitzer}\xspace}

\begin{document}

\title{CHEOPS Precision Phase Curve of the Super-Earth 55 Cnc e}

\author{B.M. Morris$^{1}$
\and L. Delrez$^{2,3,4}$
\and A. Brandeker$^{5}$
\and A. C. Cameron$^{6}$
\and A. E. Simon$^{7}$
\and D. Futyan$^{4}$
\and G. Olofsson$^{5}$
\and S. Hoyer$^{8}$
\and A. Fortier$^{7,1}$
\and B.-O. Demory$^{1}$
\and M. Lendl$^{4}$
\and T. G. Wilson$^{6}$
\and M. Oshagh$^{9,11}$
\and K. Heng$^{1}$
\and D. Ehrenreich$^{4}$
\and S. Sulis$^{8}$
\and Y. Alibert$^{7}$
\and R. Alonso$^{9,11}$
\and G. Anglada Escudé$^{12,13}$
\and D. Barrado$^{14}$
\and S. C. C. Barros$^{15,16}$
\and W. Baumjohann$^{17}$
\and M. Beck$^{4}$
\and T. Beck$^{7}$
\and A. Bekkelien$^{4}$
\and W. Benz$^{7,1}$
\and M. Bergomi$^{31}$
\and N. Billot$^{4}$
\and X. Bonfils$^{18}$
\and V. Bourrier$^{4}$
\and C. Broeg$^{7,1}$
\and T. Bárczy$^{19}$
\and J. Cabrera$^{20}$
\and S. Charnoz$^{21}$
\and M. B. Davies$^{22}$
\and D. De Miguel Ferreras$^{45}$
\and M. Deleuil$^{8}$
\and A. Deline$^{4}$
\and O. D. S. Demangeon$^{15,16}$
\and A. Erikson$^{20}$
\and H.G. Floren$^{5}$
\and L. Fossati$^{17}$
\and M. Fridlund$^{23,24}$
\and D. Gandolfi$^{25}$
\and A. Garc\'ia Mu\~noz$^{26}$
\and M. Gillon$^{2}$
\and M. Guedel$^{27}$
\and P. Guterman$^{8, 49}$
\and K. Isaak$^{46}$
\and L. Kiss$^{28,47,48}$
\and J. Laskar$^{29}$
\and A. Lecavelier des Etangs$^{30}$
\and M. Lieder$^{20}$
\and C. Lovis$^{4}$
\and D. Magrin$^{31}$
\and P. F. L. Maxted$^{32}$
\and V. Nascimbeni$^{31}$
\and R. Ottensamer$^{33}$
\and I. Pagano$^{34}$
\and E. Pallé$^{9,11}$
\and G. Peter$^{35}$
\and G. Piotto$^{31,36}$
\and A. Pizarro Rubio$^{45}$
\and D. Pollacco$^{37}$
\and F. J. Pozuelos$^{2,3}$
\and D. Queloz$^{4,38}$
\and R. Ragazzoni$^{31,36}$
\and N. Rando$^{39}$
\and H. Rauer$^{20,26,40}$
\and I. Ribas$^{12,13}$
\and N. C. Santos$^{15,16}$
\and G. Scandariato$^{34}$
\and A. M. S. Smith$^{20}$
\and S. G. Sousa$^{15}$
\and M. Steller$^{17}$
\and Gy. M. Szab{\'o}$^{41,42}$
\and D. Ségransan$^{4}$
\and N. Thomas$^{7}$
\and S. Udry$^{4}$
\and B. Ulmer$^{44}$
\and V. Van Grootel$^{3}$
\and N. A. Walton$^{43}$}

\institute{$^{1}$ Center for Space and Habitability, Gesellsschaftstrasse 6, 3012 Bern, Switzerland\\
$^{2}$ Astrobiology Research Unit, Universit\'e de Li\`ege, All\'ee du 6 Ao\^ut 19C, B-4000 Li\`ege, Belgium\\
$^{3}$ Space sciences, Technologies and Astrophysics Research (STAR) Institute, Universit{\'e} de Li{\`e}ge, All{\'e}e du 6 Ao{\^u}t 19C, 4000 Li{\`e}ge, Belgium\\
$^{4}$ Observatoire Astronomique de l'Universit\'e de Gen\`eve, Chemin Pegasi 51, Versoix, Switzerland\\
$^{5}$ Department of Astronomy, Stockholm University, AlbaNova University Center, 10691 Stockholm, Sweden\\
$^{6}$ Centre for Exoplanet Science, SUPA School of Physics and Astronomy, University of St Andrews, North Haugh, St Andrews KY16 9SS, UK\\
$^{7}$ Physikalisches Institut, University of Bern, Gesellsschaftstrasse 6, 3012 Bern, Switzerland\\
$^{8}$ Aix Marseille Univ, CNRS, CNES, LAM, Marseille, France\\
$^{9}$ Instituto de Astrof\'\i sica de Canarias, 38200 La Laguna, Tenerife, Spain\\
$^{10}$ LAM, Marseille, France\\
$^{11}$ Departamento de Astrof\'\i sica, Universidad de La Laguna, 38206 La Laguna, Tenerife, Spain\\
$^{12}$ Institut de Ci\`encies de l'Espai (ICE, CSIC), Campus UAB, Can Magrans s/n, 08193 Bellaterra, Spain\\
$^{13}$ Institut d'Estudis Espacials de Catalunya (IEEC), 08034 Barcelona, Spain\\
$^{14}$ Depto. de Astrofísica, Centro de Astrobiologia (CSIC-INTA), ESAC campus, 28692 Villanueva de la Cãda (Madrid), Spain\\
$^{15}$ Instituto de Astrof\'isica e Ci\^encias do Espa\c{c}o, Universidade do Porto, CAUP, Rua das Estrelas, 4150-762 Porto, Portugal\\
$^{16}$ Departamento de F\'isica e Astronomia, Faculdade de Ci\^encias, Universidade do Porto, Rua do Campo Alegre, 4169-007 Porto, Portugal\\
$^{17}$ Space Research Institute, Austrian Academy of Sciences, Schmiedlstrasse 6, A-8042 Graz, Austria\\
$^{18}$ Université Grenoble Alpes, CNRS, IPAG, 38000 Grenoble, France\\
$^{19}$ Admatis, Miskok, Hungary\\
$^{20}$ Institute of Planetary Research, German Aerospace Center (DLR), Rutherfordstrasse 2, 12489 Berlin, Germany\\
$^{21}$ Université de Paris, Institut de physique du globe de Paris, CNRS, F-75005 Paris, France\\
$^{22}$ Lund Observatory, Dept. of Astronomy and Theoretical Physics, Lund University, Box 43, 22100 Lund, Sweden\\
$^{23}$ Leiden Observatory, University of Leiden, PO Box 9513, 2300 RA Leiden, The Netherlands\\
$^{24}$ Department of Space, Earth and Environment, Chalmers University of Technology, Onsala Space Observatory, 43992 Onsala, Sweden\\
$^{25}$ Dipartimento di Fisica, Universit\`a degli Studi di Torino, via Pietro Giuria 1, I-10125, Torino, Italy\\
$^{26}$ Center for Astronomy and Astrophysics, Technical University Berlin, Hardenberstrasse 36, 10623 Berlin, Germany\\
$^{27}$ University of Vienna, Department of Astrophysics, Türkenschanzstrasse 17, 1180 Vienna, Austria\\
$^{28}$ Konkoly Observatory, Research Centre for Astronomy and Earth Sciences, 1121 Budapest, Konkoly Thege Miklós út 15-17, Hungary\\
$^{29}$ IMCCE, UMR8028 CNRS, Observatoire de Paris, PSL Univ., Sorbonne Univ., 77 av. Denfert-Rochereau, 75014 Paris, France\\
$^{30}$ Institut d'astrophysique de Paris, UMR7095 CNRS, Université Pierre \& Marie Curie, 98bis blvd. Arago, 75014 Paris, France\\
$^{31}$ INAF, Osservatorio Astronomico di Padova, Vicolo dell'Osservatorio 5, 35122 Padova, Italy\\
$^{32}$ Astrophysics Group, Keele University, Staffordshire, ST5 5BG, United Kingdom\\
$^{33}$ Department of Astrophysics, University of Vienna, Tuerkenschanzstrasse 17, 1180 Vienna, Austria\\
$^{34}$ INAF, Osservatorio Astrofisico di Catania, Via S. Sofia 78, 95123 Catania, Italy\\
$^{35}$ Institute of Optical Sensor Systems, German Aerospace Center (DLR), Rutherfordstrasse 2, 12489 Berlin, Germany\\
$^{36}$ Dipartimento di Fisica e Astronomia "Galileo Galilei", Università degli Studi di Padova, Vicolo dell'Osservatorio 3, 35122 Padova, Italy\\
$^{37}$ Department of Physics, University of Warwick, Gibbet Hill Road, Coventry CV4 7AL, United Kingdom\\
$^{38}$ Cavendish Laboratory, JJ Thomson Avenue, Cambridge CB3 0HE, UK\\
$^{39}$ ESTEC, European Space Agency, 2201AZ, Noordwijk, NL\\
$^{40}$ Institut für Geologische Wissenschaften, Freie Universität Berlin, 12249 Berlin, Germany\\
$^{41}$ ELTE Eötvös Loránd University, Gothard Astrophysical Observatory, 9700 Szombathely, Szent Imre h. u. 112, Hungary\\
$^{42}$ MTA-ELTE Exoplanet Research Group, 9700 Szombathely, Szent Imre h. u. 112, Hungary\\
$^{43}$ Institute of Astronomy, University of Cambridge, Madingley Road, Cambridge, CB3 0HA, United Kingdom\\
$^{44}$ Ingenieurbüro Ulmer - Technische Informatik, Im Technologiepark 1, 15236 Frankfurt, Germany\\
$^{45}$ Airbus DS Spain\\
$^{46}$ Science and Operations Department - Science Division (SCI-SC), Directorate of Science, European Space Agency (ESA), European Space Research and Technology Centre (ESTEC)\\
$^{47}$ ELTE E\"otv\"os Lor\'and University, Institute of Physics, P\'azm\'any P\'eter s\'et\'any 1/A, 1117 Budapest, Hungary\\
$^{48}$ Sydney Institute for Astronomy, School of Physics A29, University of Sydney, NSW 2006, Australia\\
$^{49}$ Division Technique INSU, BP 330, 83507 La Seyne cedex, France}

   \date{Received 2021}
 
  \abstract
   {55 Cnc e is a transiting super-Earth (radius $1.88\rm\,R_\oplus$ and mass $8\rm\, M_\oplus$) orbiting a G8V host star on a 17-hour orbit. \spitzer observations of the planet's phase curve at 4.5 $\mu$m revealed a time-varying occultation depth, and MOST optical observations are consistent with a time-varying phase curve amplitude and phase offset of maximum light. Both broadband and high-resolution spectroscopic analyses are consistent with either a high mean molecular weight atmosphere or no atmosphere for planet e. A long term photometric monitoring campaign on an independent optical telescope is needed to probe the variability in this system.}
   {We seek to measure the phase variations of 55 Cnc e with a broadband optical filter with the 30 cm effective aperture space telescope CHEOPS and explore how the precision photometry narrows down the range of possible scenarios.}
   {We observed 55 Cnc for 1.6 orbital phases in March of 2020. We designed a phase curve detrending toolkit for CHEOPS photometry which allows us to study the underlying flux variations of the 55 Cnc system.}
   {We detected a phase variation with a full-amplitude of $72 \pm 7$ ppm but do not detect a significant secondary eclipse of the planet. The shape of the phase variation resembles that of a piecewise-Lambertian, however the non-detection of the planetary secondary eclipse, and the large amplitude of the variations exclude reflection from the planetary surface as a possible origin of the observed phase variations. They are also likely incompatible with magnetospheric interactions between the star and planet but may imply that circumplanetary or circumstellar material modulate the flux of the system.}
   {Further precision photometry of 55 Cnc from CHEOPS will measure variations in the phase curve amplitude and shape over time this year.}


   \maketitle

\section{Introduction}

55 Cnc e is perhaps the rocky planet most amenable to characterization with CHEOPS. This super-Earth with radius $1.88\rm\,R_\oplus$ and mass $8\rm\, M_\oplus$ orbits a very bright ($V=6$) G8V host star \citep{Demory2011, Bourrier2018}. The planet was initially detected with an apparent $2.8$-day orbital period \citep{McArthur2004,Fischer2008}; this was later revised to the 17-hour period \citep{Dawson2010} which enabled the detection of the transit in the following year \citep{Winn2011}.

There is evidence for two distinct plausible climatic scenarios for 55 Cnc e: either it is a lava world without a significant atmosphere \citep{Demory2016b}, or it has a very thick atmosphere \citep{Angelo2017, Bourrier2018}. Existing transit observations of 55 Cnc e at 3.6 and $4.5\,\mu$m show similar transit depths, also consistent with an opaque atmosphere or no atmosphere. \citet{Mahapatra2017} studied cloud properties and found that mineral clouds could form in a thick atmosphere scenario for 55 Cnc e.

Ultra-short period rocky planets may not have been born that way. A growing body of evidence suggests that some ultra-short period planets may have once been giants that lost most of their gaseous envelopes due to photoevaporation \citep{Lecavelier2004, Owen2017, Fulton2017, VanEylen2018}. 55 Cnc e may represent one end-point of planetary evolution under high irradiation. 

\subsection{Phase curve}

Though the planet's density is consistent with a purely rocky composition \citep{Madhusudhan2012,Dorn2017}, its \spitzer $4.5\,\mu$m phase curve peaks prior to the secondary eclipse, and indicates a large day-night temperature contrast \citep{Demory2016b}. However, \citet{Kite2016} and \citet{Angelo2017} have argued that the $4.5\,\mu$m phase curve indicates that the day-to-night side heat redistribution is too efficient to be explained by heat transport due to currents in a molten lava ocean. They suggest that the $4.5\,\mu$m phase curve observations may instead require the presence of a thick atmosphere with pressure $\gtrsim 1.4$ bars and temperature 2400~K at the photosphere. 

Recently, \citet{Sulis2019} used MOST photometry to measure variability in the phase curve of 55 Cnc e, the amplitude and phase of maximum light are seen to vary from year to year from 113 to 28 ppm and from 5 to 217 degrees. The authors were not able to identify a single origin of this variability. Different scenarios were proposed: star-planet interactions, the presence of a transiting circumstellar torus of dust, or an instrumental artifact. 

\subsection{Secondary Eclipse}

The secondary eclipse depth of the planet at $4.5\,\mu$m varied by a factor of three between 2012 and 2013 \citep{Demory2016a, Tamburo2018}. The corresponding thermal emission from the dayside atmosphere must have varied by $\sim$1500 K between the two epochs, which could be explained by atmospheric albedo variability, or large-scale surface volcanic activity, for example, like the volcanic plumes observed on Jupiter's moon Io \citep[e.g.][]{Spencer1997, McEwen1998}. These dynamic features were detected at $4\sigma$ significance \citep{Demory2016a}. The secondary eclipse was not detected with MOST \citep{Sulis2019}, and a weak $3\sigma$ secondary eclipse has been claimed from TESS observations \citep{Kipping2020}.

\subsection{Transit}

Previous observations support the presence of an atmosphere with a lack of H$_2$O and a hint of HCN \citep{Tsiaras2016, Esteves2017}, despite the lack of H detection by \citet{Ehrenreich2012}. It is possible that an H$_2$-rich atmosphere could be present without a detection of H, as the H may get swept away by radiation pressure. However the lack of Ly-$\alpha$ and helium absorption from 55 Cnc e may point to a lack of a thick modern atmosphere \citep{Zhang2021}. This alternatively may imply a C/O ratio greater than solar \citep{Madhusudhan2012}. In this regime, carbon- and nitrogen-based molecules such as HCN, N$_2$ and CO are expected to be the trace gasses in the atmosphere \citep{Venot2015, Miguel2019}. In such an environment, one might also expect CH$_4$, C$_2$H$_2$, and others to be detected \citep{Heng2016}. The \spitzer $4.5 \,\mu$m transits, however, yield a radius consistent with optical and near-IR observations, implying a flat transmission spectrum \citep{Demory2016a}. 

\subsection{Host star}

Stellar magnetic activity is also a possible culprit in the apparent phase variations. The rotation period of the star is approximately 40 days \citep{Henry2000,Fischer2008}. \citet{Folsom2020} mapped the large-scale magnetic field of 55 Cnc and find a 5.8 G dipole, tilted with respect to the rotation axis. Simulations of the stellar wind predict that planet e orbits inside the Alfv{\'e}n radius, so the planet's influence on the stellar wind propagates back to the stellar surface. This could result in star-planet interactions. On long timescales, \citet{Bourrier2018} used photometric and spectroscopic observations to constrain the magnetic activity cycle of the host star $P_\mathrm{cyc} = 10.47 \pm 0.21$ years. Spectroscopic observations on the order of weeks may be required to constrain the relevant timescales for stellar magnetic activity. Far-UV observations of 55 Cnc with HST also showed chromospheric variability at many timescales, which may be linked to star-planet interactions as well \citep{Bourrier2018b}.

\subsection{Outline}

In this work, we aim to probe the following questions with CHEOPS photometry: (1) is reflected light a plausible source of the photometric variability, (2) is stellar variability on the orbital period of the planet an equally compatible source, (3) what is the limiting precision of CHEOPS photometry for bright stars, (4) do we detect a significant secondary eclipse in the optical? In Section~\ref{sec:cheops} we give background on the CHEOPS telescope and mission design, and the details of the observations presented in this work. In Section~\ref{sec:systematics} we outline methods for mitigating systematics in CHEOPS observations like those of 55 Cnc e. In Section~\ref{sec:phot} we devise a photometric model for the detrended CHEOPS photometry of 55 Cnc e. We discuss the interpretation of the results in Section~\ref{sec:discussion} and conclude in Section~\ref{sec:conclusion}.

\section{CHEOPS} \label{sec:cheops}

\subsection{Background} \label{sec:cheops_background}

CHEOPS is an on-axis Ritchey-Chr{\'e}tien telescope with a 320 mm diameter primary mirror \citep{Benz2020}, optimized for high-precision photometry \citep{Lendl2020}. The primary mirror is partially obscured by the secondary mirror, which is held in place by three supports. The telescope is intentionally defocused, and due to the secondary mirror supports, CHEOPS exposures have a distinctive three-pointed point-spread function (PSF; see Figure~\ref{fig:smear_correction}), with its core spanning 32 pixels across. 

The spacecraft is nadir-locked in a low-Earth orbit about $700$ km from the Earth's surface, which causes stars in the field of view to rotate around the line of sight once per CHEOPS orbit. Neighbouring stars can introduce periodic photometric systematics, that we will discuss in Sections~\ref{sec:roll_angle}
and \ref{sec:smearing}.

The CHEOPS spacecraft always points its backside towards the Sun, and variable spacecraft heating may occur when slewing between targets. Earth occultations of the target star also cause temperature perturbations in the telescope optical assembly. We will discuss mitigation for temperature variations during a measurement in Section~\ref{sec:ramp}. 

CHEOPS exposures are recorded by a back-illuminated, frame-transfer CCD, which spans a broad, optical bandpass similar to the Gaia $G$ filter. Exposures less than 23 seconds in duration are stacked on-board to save downlink bandwidth (the image downlink cadence varies between 23 and 60 seconds depending on the exposure time). However, before stacking,  ``imagettes'' of 30 pixels in radius are extracted. These small images are centered in the CHEOPS field of view and contain the PSF of the target star. In the case of the 55 Cnc observations, imagettes were stacked in pairs  before being downlinked to the ground. As a result of these frequent reads (the cadence of each stacked imagette is 4.4 seconds) and the well-resolved PSF of CHEOPS, it is possible to do precise PSF photometry on the target star -- we will discuss insights on the 55 Cnc e photometry from PSF photometry in Section~\ref{sec:psf}.

\subsection{Observations}

A single primary transit of 55\,Cnc\,e was observed during the CHEOPS In-Orbit Commissioning (IOC) phase on UT 2020 March 09. Due to the brightness of the target, an effective exposure time of 30.8\,s was achieved by the stacking of 14 individual readouts of 2.2\,s. Imagettes were stacked on board in pairs, which means that with a cadence of 30.8\,s seconds a stacked image and 7 stacked imagettes were downlinked. CHEOPS also observed 1.6 orbital phases of 55\,Cnc over 26\,hours on UT 2020 March 23-24, stacking 20$\times$2.2\,s individual frames and thus obtaining a final 44.5\,s cadence. The log of the observations is presented in Table\,\ref{tab:obs_log}. Gaps in the observations correspond to occultations of the target by the Earth, and passes by the spacecraft through the South Atlantic Anomaly, during which images are discarded as there is a significantly increase in the cosmic ray hits. 

The observations were automatically reduced by the CHEOPS Data Reduction Pipeline \citep[DRP v12,][]{Hoyer2020}. The DRP takes the raw images from the spacecraft and produces time-series photometry of the target star for analysis by the user. The DRP produces four light curve data products for each visit, called the ``DEFAULT, RSUP, RINF'' and ``OPTIMAL'' apertures. In the following analysis, we focus on the ``DEFAULT'' aperture product with a circular aperture of 25 pixels in radius.

\begin{table*}[]
    \centering
    \begin{tabular}{ccccccc}
    \hline
    Date Start & Date Stop & File Key & Duration & Exposure  & Exposures  & Efficiency \\
    $[$UT]& [UT] & & [hh:mm] & Time [s] & per stack & \%\\ 
    \hline
    2020-03-09 04:55 & 2020-03-09 11:15 & {\small CH\_PR300024\_TG000301} & 06:14 & 30.8 & 14 ($\times$2.2\,s) & 59\\
    2020-03-23 13:40 & 2020-03-24 16:01 & {\small CH\_PR100041\_TG000601} & 26:15 & 44.0 & 20 ($\times$2.2\,s) & 56 \\
    \hline
    \end{tabular}
    \caption{CHEOPS observation logs, corresponding to the IOC observations in the first row and the phase curve observations in the second row. The File Key is useful for uniquely identifying the visits used in this work.}
    \label{tab:obs_log}
\end{table*}

\section{Mitigating CHEOPS Spacecraft Systematics} \label{sec:systematics}

The design of the CHEOPS spacecraft introduces several systematics to precision photometry which we seek to outline and account for below. 

\begin{figure*}
    \centering
    \includegraphics{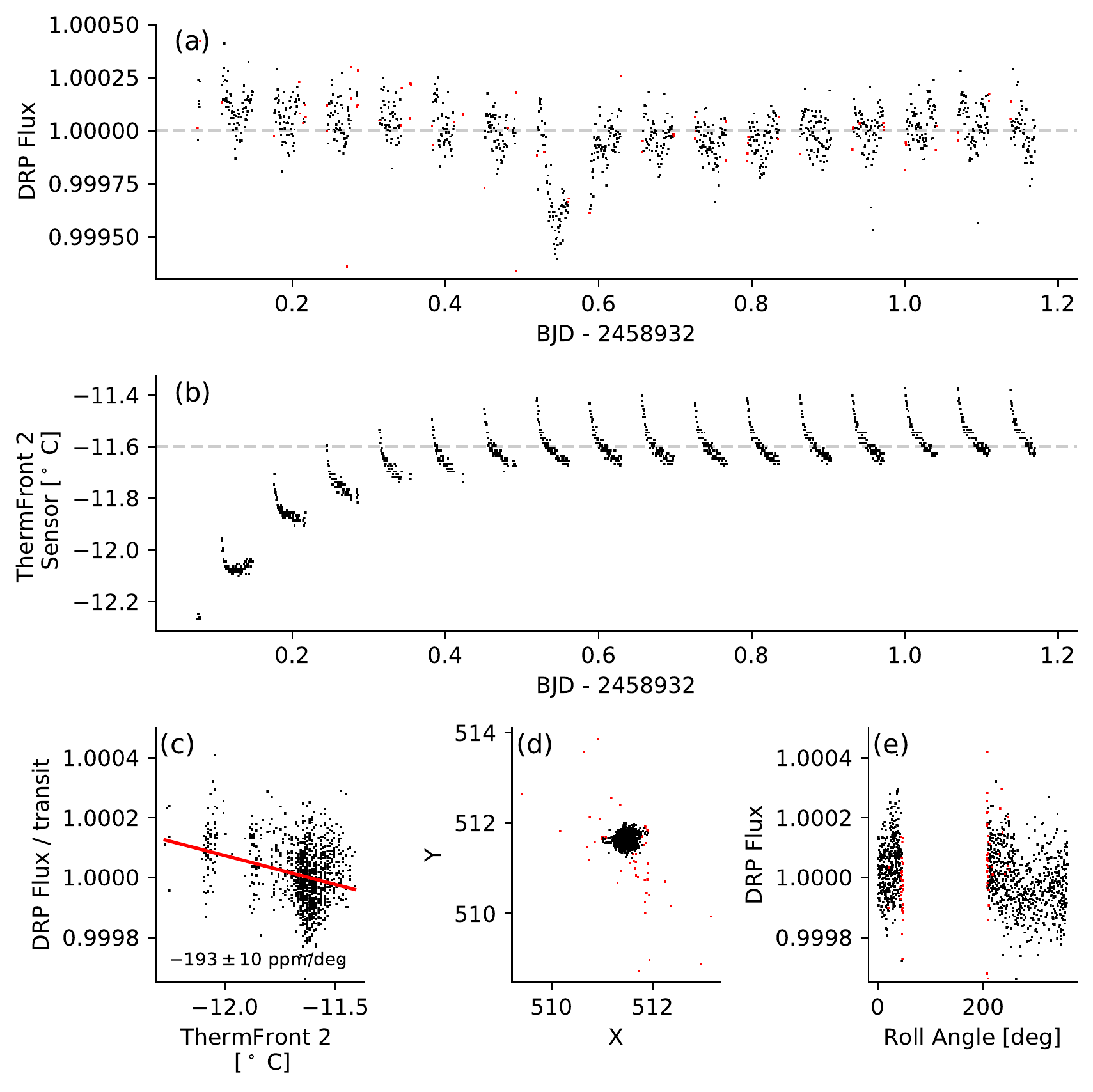}
    \caption{Photometry from CHEOPS of 55 Cnc e and detrending vectors. (\textbf{a}) Raw fluxes returned by the Data Reduction Pipeline (DRP) aperture photometry observations over the 1.6 orbital phases of 55 Cnc e, centered on a transit. Red points are centroid outliers and are masked from analysis. The gaps in the observations occur primarily when Earth is occulting the target star. The horizontal dashed line at flux unity represents the median flux, about which we see variations in flux in the phase curve. (\textbf{b}) One component of the variations in the raw phase curve in panel (a) is a ``ramp'' feature which is anticorrelated with the telescope tube temperature, denoted the ``ThermFront 2 Sensor'' temperature. A full discussion of the link between the flux variations and telescope tube temperature is written in Section~\ref{sec:ramp}. The temperature of the telescope tube rises asymptotically from a minimum at the beginning of the observations to a near-constant value after $\sim 0.5$ days. There is an additional perturbation in the temperature once per orbit, thought to be caused by illumination of the telescope optical assembly by the Earth during occultation. (\textbf{c}) Changes in the DRP flux are associated with changes in the ThermFront 2 Sensor temperature after dividing out the transit with a slope of  $-193 \pm 10$ ppm/degree Celsius. (\textbf{d}) Measured centroid of the target star on the detector in pixel units -- outliers in red are masked from further analysis. (\textbf{e}) Trends in DRP flux with roll angle after correction for smearing (see Section~\ref{sec:smearing}) are $\lesssim 100$ ppm, and are a result of correlations between the spacecraft roll angle and the contamination from nearby stars, and in the background light (see Section~\ref{sec:roll_angle}). Red points are the masked centroid outliers, which mostly occur immediately before and after the Earth occultation.} 
    \label{fig:raw}
\end{figure*}

\subsection{Pointing}

The spacecraft pointing is generally better than 1 arcsecond, where 1 arcsecond spans about one pixel \citep{Benz2020}. However, the exposures that occur just before and after Earth occultation are often more likely to show significant pointing offsets -- see Figure~\ref{fig:raw}d. We mask out all photometry with centroids $>3.5\sigma$ away from the median centroid.

\subsection{Roll Angle} \label{sec:roll_angle}

Due to the nadir-locked orientation of CHEOPS, the field of view of the telescope rotates once per CHEOPS orbit. In practice, this means that the field stars rotate about the target star. The core of the PSF is 32 pixels across, but it has extended wings out to $\sim 50$ pixels. The extended and irregular shape of the CHEOPS PSF and the rotation of the field, leads to variations in the flux inside the photometric aperture which is centered on the target star. The sky background flux is also a strong function of the position of CHEOPS in its orbit, as background light from Earthshine contaminates the aperture close to Earth occultation.
While these sources of contamination are present in all CHEOPS observations, in the case of 55 Cnc they do not play a major role due to the brightness of the star. However, decorrelating the flux of the target star against the roll angle of the spacecraft is necessary to ensure the rotating field does not introduce correlated noise in the photometry.

\subsubsection{Smearing} \label{sec:smearing}

\begin{figure}
    \centering
    \includegraphics[width=0.95\linewidth]{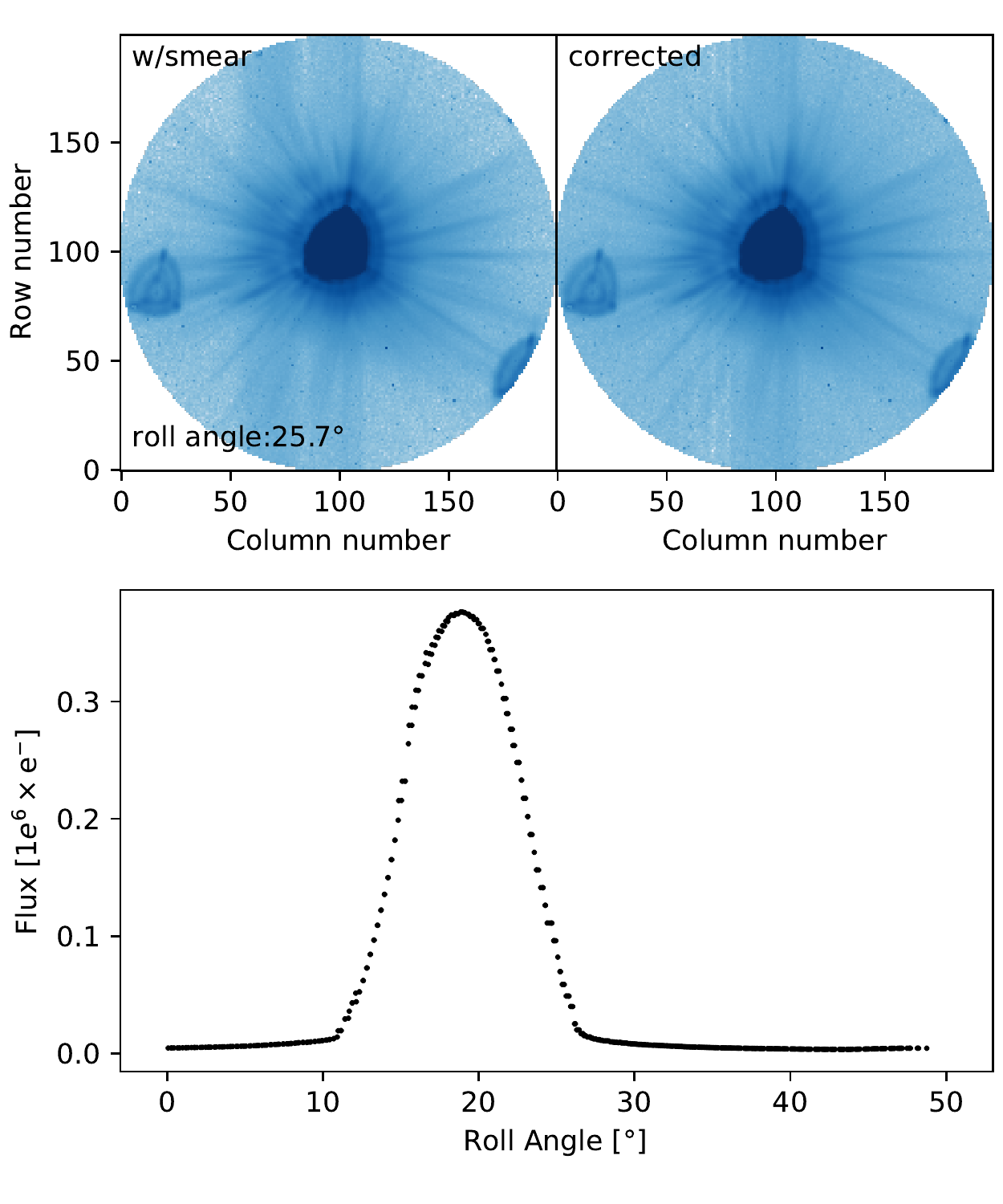}
    \caption{Smear contamination correction. {\it Top-left}: an example CHEOPS exposure where the smear contamination from 53 Cnc, located outside the sub-array image, is visible at the left of 55 Cnc (the star at the center of the field of view). {\it Top-right}: the same frame after the DRP v12 smear correction.  {\it Bottom}: the estimated smear contamination flux within the DEFAULT aperture as a function of the observation roll angle which affects each orbit. }
    \label{fig:smear_correction}
\end{figure}

CHEOPS has no shutter to block the incoming light between one exposure and the next. This means that light continues to be collected during the frame transfer from the exposed section to the covered, read-out section. It takes 25 milliseconds to transfer the full CCD image area (1024 pixel-rows) to the covered section for read-out. Therefore, a consequence of this constant transfer speed is the presence of ``smearing trails'' in the images. Smearing trails are vertical stripes of flux going through every PSF in an image. All stars in the CHEOPS field of view (full CCD) leave a smearing trail after each exposure. Because the transfer time is constant, i.e  independent of the exposure time and the stellar brightness, smear trails are stronger for brighter stars. Therefore, while in most of the cases they are hard to detect, the flux level of the trails becomes significant when the target star or a neighbour star is very bright. It is important to make the distinction between the self-smear trail of the target star and the smearing trails from the field stars. The smearing trail of the target star is always ``fixed'' in the image, only following the small jitter of the target. The photon noise is negligible and it has no impact in the photometry as it almost does not change during the whole visit. However, things are much more complex with the smearing trails of the field stars. As the neighbouring stars rotate around the target, the vertical smearing trails sweep back and forth over the target star (on the horizontal axis), producing a highly predictable positive flux anomaly in uncorrected photometric extractions. Any star in the field of view (within a radius of 17 arcmin) may contaminate the photometric aperture with its smearing trails at least once per orbit. Figure~\ref{fig:smear_correction} top-left and bottom illustrate the problem in the case of 55 Cnc and the bright neighbour 53 Cnc (Gaia $G$ magnitude 5.17).

As of DRP v12 \citep{Hoyer2020}, the effect of smearing is removed from each sub-array frame before photometry is computed (see Figure~\ref{fig:smear_correction} top-right). This is accomplished by forward modeling the the smearing produced by each star in CHEOPS CCD, except for the target, to estimate the flux deviations due to the smearing. Finally, this estimated smear signal is directly subtracted from each sub-array frame. No further smear corrections are needed in the photometric analysis that follows  -- see Figure~\ref{fig:raw}e. 

\subsection{The ``Ramp''} \label{sec:ramp}

Space-based photometry is often affected by systematic variations in the measured flux from quiescent target stars, sometimes called ``ramps''. Ramps have been observed in photometry from the Hubble Space Telescope (HST) WFC3 \citep[e.g.:][]{Berta2012, Zhou2017, Stevenson2019}, HST/STIS \citep{Demory2015}, and Spitzer/IRAC \citep[e.g.:][]{Deming2006, Deming2009, Desert2009, Knutson2008, Demory2011}. Charge trapping is one proposed mechanism for producing the observed ramps in WFC3 and IRAC photometry \citep[e.g.:][]{Agol2010, Zhou2017}.

A ramp feature has been observed in CHEOPS photometry on some targets including 55 Cnc -- see Figure~\ref{fig:raw}a. The amplitude of the ramp feature in the 55 Cnc e phase curve time series is $\sim 100$ ppm, showing a decrease in the apparent flux of the target star over a span of $\sim 0.5$ days at the beginning of the visit. The ramp effect has been observed on other targets as well, sometimes as an asymptotically increasing flux trend, and sometimes as a decreasing one.

Our current best hypothesis for the cause of the ramp feature in CHEOPS photometry is solar thermal forcing of the telescope optical assembly. The scale of the PSF has been noted to vary slightly with time, most sharply at the beginning of a time series, and settles to a constant scale factor after several hours (more on PSF photometry in Section~\ref{sec:psf}). The changing scale of the PSF could be due to small differences in the distance between the primary and secondary mirrors as a result of temperature changes in the telescope tube as the solar illumination on the spacecraft varies between pointings. The change in PSF scale with time may modulate the flux which falls within fixed apertures, causing the apparent ramps in flux. 

If the observed flux ramp is the result of changes in the PSF scale caused by thermal settling of the spacecraft, then we can remove the ramp by detrending against the temperature of the telescope tube -- see Figure~\ref{fig:raw}c. The telescope temperature is recorded in the housekeeping time series from the spacecraft, shown in Figure~\ref{fig:raw}b. We include this temperature time series as a basis vector in the detrending analysis in Section~\ref{sec:phot}, which removes the $-193 \pm 10$ ppm variation in flux per degree Celsius in the DRP photometry.

\subsection{Insights from PSF photometry} \label{sec:psf}

The defocused PSF is sampled over $\sim 32$ pixels, allowing us to build up a precise PSF model which can be fit to each exposure via \texttt{PIPE}: the PSF Imagette Photometric Extraction pipeline. PSF photometry has a few distinct benefits for validating detections made with the standard DRP (aperture) photometry, since PSF photometry: (1) is resilient against cosmic ray strikes, bad pixels, and detector artifacts such as smearing, because it fits a model to the whole of the PSF rather than simply summing within an aperture, (2) explicitly accounts for changes in the scale of the PSF as a function of time, and (3) gives access to the sub-exposures that make up each exposure stack, and thus has finer time sampling than DRP photometry. We enumerate details on the PSF photometric extraction in Appendix~\ref{sec:psf_appendix}, including a direct comparison of the DRP and PSF photometry in Figure~\ref{fig:psf_drp}. There are a few challenges associated with PSF photometry which affect DRP photometry to a lesser extent, including for example measuring the background from the small imagettes and source blurring due to pointing jitter during the exposure, so we present results with both the DRP and PSF photometry in this work. 

\section{Detrended CHEOPS Photometry of 55 Cnc} \label{sec:phot}

In this section, we aim to measure the shape and amplitude of the variations in flux of the 55 Cnc system while simultaneously accounting for: the systematics which affect the photometry, the transit, and any evidence of planetary occultations.

In Section~\ref{sec:basisvectors}, we select a small set of basis vectors which we will use to detrend the light curve. In Section~\ref{sec:transit}, we detail the model used to fit the two transits. We outline a series of models for the observed phase variations in Section~\ref{sec:models}, each with the same set of basis vectors but with a unique, physically-motivated phase variation model. We measure the weights and uncertainties on each set of basis vectors and the phase variation model with the No U-Turn Sampler (NUTS) implemented by \texttt{PyMC3} \citep{pymc3}, and compare the predictive power of each model with Leave-One-Out Cross-Validation \citep[LOO-CV;][]{Vehtari2015}.

\subsection{Detrending basis vectors} \label{sec:basisvectors}

We will detrend the light curve against a set of basis vectors which are correlated with the flux of 55 Cnc. There are many ``housekeeping'' sensor readouts, as well as photometric by-products like target centroids and background levels, which are packaged in the DRP data products as time series along with the times and fluxes. 

We performed an exhaustive search through these candidate detrending basis vectors to identify which vectors yield the most significant improvements to the light curve fits without overfitting. We quantify the improvement of each candidate basis vector in the detrended light curve with the Bayesian Information Criterion (BIC), and find that a combination of four basis vectors produce the minimum (ideal) BIC. These four vectors are: the cosine and sine of the spacecraft roll angle (see description of roll angle in Section~\ref{sec:roll_angle}), the ThermFront 2 thermistor readout which tracks the telescope temperature (see description in Section~\ref{sec:ramp}), and a unit vector. The unit vector accounts for the constant mean flux of the star. Adding further basis vectors such as the PSF centroid or higher order functions of the earlier basis vectors are disfavored by the BIC.

\subsection{Transit and eclipse} \label{sec:transit}

We simultaneously fit the two transits of 55 Cnc e along with the detrending basis vectors and the phase curve models with an \citet{exoplanet:agol20} transit model assuming the period from \citet{Bourrier2018}, and the interferometric stellar radius and derived mass by \citet{vonBraun2011}. We fit for quadratic limb-darkening parameters using the efficient triangular sampling method of \citet{Kipping2013}. The secondary eclipse of the planet is fixed in time to occur at the expected eclipse time for zero eccentricity, consistent with the small or zero eccentricity found by \citet{Bourrier2018}.

\begin{figure*}
    \centering
    \includegraphics[width=\textwidth]{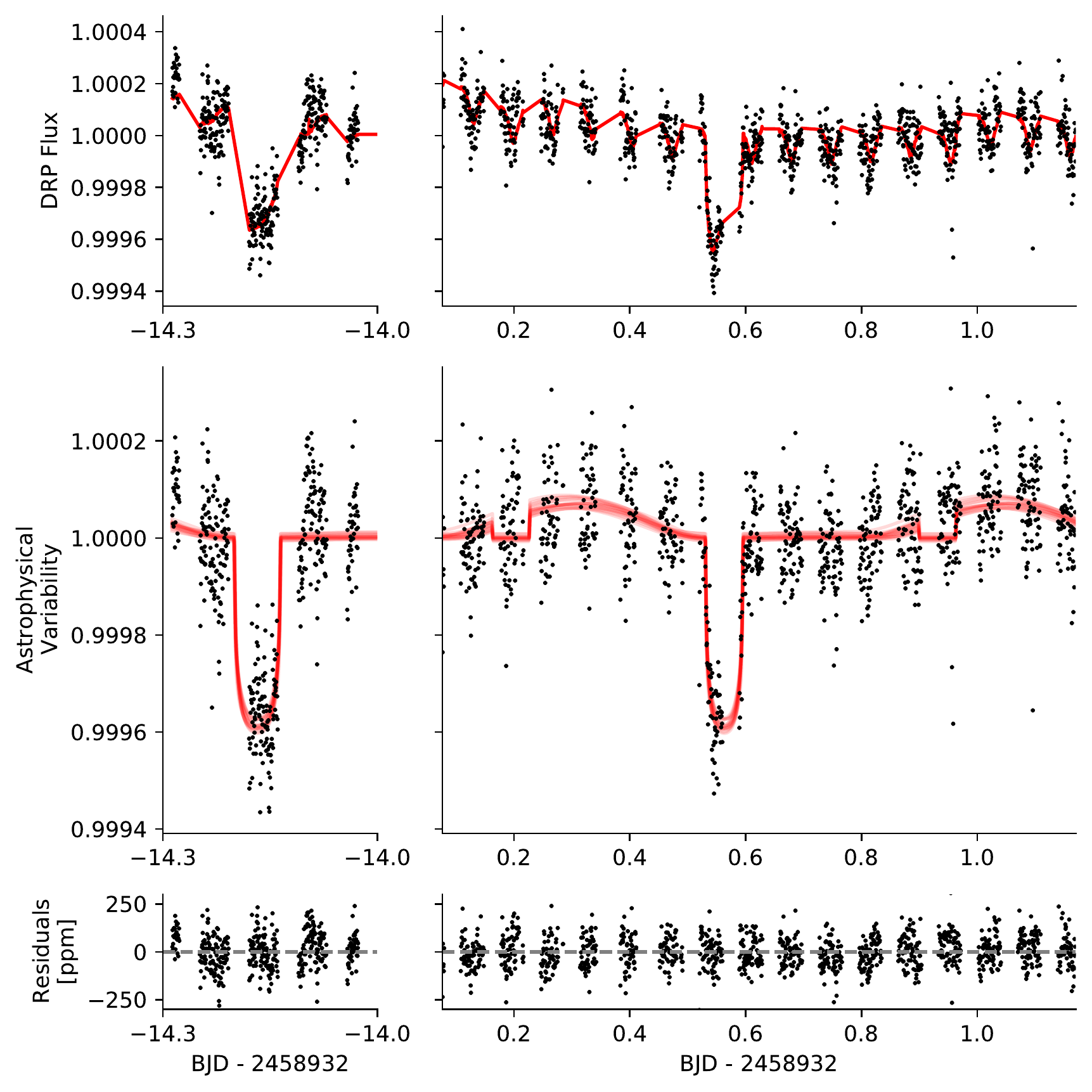}
    \caption{Detrended CHEOPS photometry of 55 Cnc e. {\bf Upper}: The Data Reduction Pipeline (DRP) aperture photometry fluxes in black, and in red the best-fit linear combination of a transit model, a piecewise-Lambertian phase variation, and several detrending vectors. This fit is used to infer the best-fit transit parameters. The left column shows transit observations obtained during the in-orbit commissioning (IOC) phase, and the right column shows the phase curve observation. {\bf Middle}: The residuals after the instrumental and systematic variations have been removed (black) and several draws from the posterior distributions for the transit and phase curve parameters (red). {\bf Lower}: The residuals after removing all systematics and astrophysical signals from the observations.}
    \label{fig:components}
\end{figure*}

\subsection{Phase variation models} \label{sec:models}

Several physically-motivated models could be applied to the out-of-transit phase variability of the 55 Cnc e system. We use: a Lambertian model, which naturally occurs when a planet emits or scatters light isotropically; the piecewise-Lambertian model of \citet{Hu2015} which describes a planetary surface that behaves like a Lambertian sphere between two longitudes; a simple sinusoidal model which assumes the flux is varying with the period of the planet and an arbitrary phase offset. We detail the sampling procedure in Section~\ref{sec:sampling} and compare the model fits in Section~\ref{sec:comparison}.

In each following section outlining a phase curve model, we describe the baseline model as: 
\begin{equation}
    f = {\bf X} \hat{\beta}
\end{equation}
which is the best-fit weights $\hat{\beta}$ to the linear combination of basis vectors in the design matrix ${\bf X}$, which is composed of the following column vectors: the cosine and sine of the roll angle, a unit vector, and the ``ThermFront 2'' sensor temperature.

\subsubsection{Sinusoidal model}

The simplest model to describe the phase variations of the 55 Cnc e system is simply cosine-shaped variability with a fixed period set to the planet's orbital period, 
\begin{equation}
f = {\bf X}\hat{\beta} + a\cos\left(\xi - \phi - \pi\right)
\end{equation}
where $\xi$ is the orbital phase, which is defined on $[-\pi, \pi]$ where $|\xi| = \pi$ is the time of transit and $\xi = 0$ is the secondary eclipse. We fit for the amplitude $a$, phase offset $\phi$. 

We do not assume that the origin of the sinusoidal signal is the planet, which would imply that the eclipse depth should be equivalent to the amplitude of the sinusoid at the mid-eclipse time, and the in-eclipse fluxes would be constant. As a result, our simple model changes in flux during the time of the expected secondary eclipse, and the eclipse depth is an unconstrained free parameter.

\subsubsection{Lambertian model}

A Lambertian sphere isotropically scatters or emits thermally, giving a phase function: 
\begin{equation}
f = {\bf X}\hat{\beta} + a E \left(\sin(|\xi|) + (\pi - |\xi|)\cos(|\xi|)\right).
\end{equation}
Here the phase function is modulated by an amplitude parameter $a$ and an eclipse function $E$ which is the eclipse model with no limb darkening, which models the eclipse of the planet. By construction, this model has no phase offset.

\subsubsection{Piecewise-Lambertian model}

Appendix A of \citet{Hu2015} defines a piecewise-Lambertian sphere for describing the Kepler optical phase curve of the apparently asymmetric planet Kepler-7 b. The model is constructed with significant reflection or emission between two longitudes, and negligible reflection/emission elsewhere. We implement the \citet{Hu2015} model to test if an asymmetric planetary albedo distribution could explain the apparently asymmetric phase variations of 55 Cnc e. This model incurs two additional fitting parameters corresponding to the longitudes $\xi_1,~\xi_2$ which bound the less reflective hemisphere.

\subsection{Posterior sampling} \label{sec:sampling}

For each of these models, we sample the posterior distributions for the transit parameters, the $\hat{\beta}$ best-fit estimators, as well as the parameters relevant to each phase curve model, using the No U-Turn Sampler (NUTS) implemented by \texttt{PyMC3} \citep{pymc3}. The NUTS is a variant of Hamiltonian Monte Carlo, which is a gradient-based Monte Carlo approach that has been shown to be  computationally efficient for sampling posterior distributions for high-dimensional problems \citep[see e.g.][]{Betancourt2017}.

We also treat the flux uncertainty on each photometric measurement as a free parameter, and introduce a term to the likelihood that penalizes larger uncertainties. This has the effect of finding the smallest uncertainties that bring the reduced $\chi^2$ to unity. The maximum-likelihood solution has a uncertainty per flux of $84 \pm 3$ ppm, which is larger than the DRP estimated flux uncertainty of 51 ppm.

\begin{figure*}
    \centering
    \includegraphics[scale=0.9]{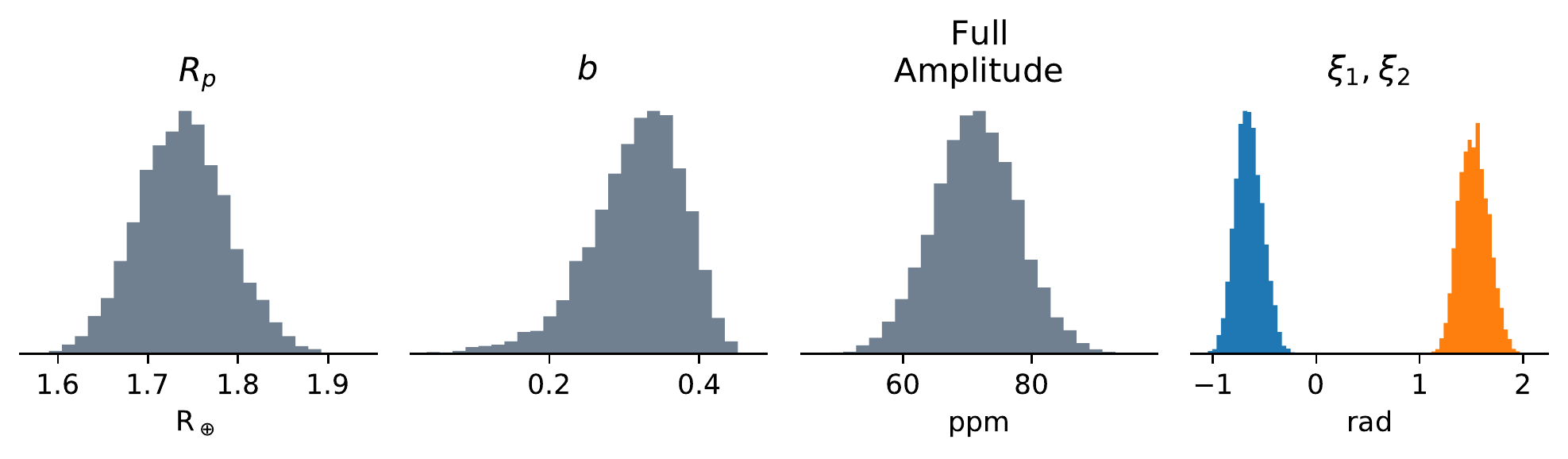}
    \caption{Select posterior distributions for key parameters in the composite transit, eclipse and piecewise-Lambertian phase curve model. The parameters are the planet radius $R_p$, impact parameteter $b$, full amplitude of the phase variations, and the start and stop longitudes of the piecewise-Lambertian model $\xi_1, \xi_2$. The full posterior distribution correlation matrix is given in Appendix~\ref{app:corner}.}
    \label{fig:posteriors}
\end{figure*}

Figure~\ref{fig:posteriors} shows posterior distributions for some key parameters, and the full joint posterior correlation matrix is shown in Appendix~\ref{app:corner}.

\subsection{Model comparison} \label{sec:comparison}

We measure the relative likelihood that each model describes the observations while penalizing the models with more free parameters with Leave-One-Out Cross-Validation statistic and the Widely Applicable Information Criterion \citep[LOO-CV and WAIC;][]{Vehtari2015}. Both techniques are a computational efficient method for estimating the relative likelihood that one model in a set is preferred over the others, using only the posterior samples from the pre-computed Markov chains used for parameter inference. The preferred models have the smallest $\Delta$LOO or $\Delta$WAIC, where more significant preferences for a given model have, for example, $\Delta$LOO$>10$. The results are given in Table~\ref{tab:loo}; the piecewise-Lambertian model is the preferred model, followed closely by the sinusoidal model, and finally the symmetric Lambertian model. We adopt the physically-motivated piecewise-Lambertian model as the most likely in the following discussion, noting that the empirical sinusoidal model fits about as well as the piecewise-Lambertian. 

\begin{table}
\centering
\begin{tabular}{l|cc|cc}
      & \multicolumn{2}{c}{LOO} & \multicolumn{2}{c}{WAIC} \\ 
Model & $\Delta$ & Weight & $\Delta$ & Weight\\ \hline
Piecewise-Lam & -- & 0.62 & -- & 0.62 \\
Sinusoid & 4 & 0.38 & 4 & 0.38 \\
Lambert & 48 & 0.00 & 48 & 0.00 \\
No var. & 57 & 0.00 & 57 & 0.00 \\
\end{tabular}
\caption{Change in the Leave-One-Out Cross-Validation and Widely Applicable Information Criterion statistics between models compared with the best-fitting model, which is the piecewise-Lambertian model of \citet{Hu2015}. The smaller the $\Delta$LOO or $\Delta$WAIC, the more compatible the data are with a given model. The weight for each statistic can be interpreted as the probability that a given model is preferred over the others. The statistic values and probabilities evaluated with both techniques are quite similar. The Piecewise-Lambertian model is similarly compatible with the sinusoidal model, but the sinusoidal model lacks a physical interpretation so we favor the Piecewise-Lambertian. The pure Lambertian and no-variation models are ruled out. \label{tab:loo}}
\end{table}

\begin{figure*}
    \centering
    \includegraphics{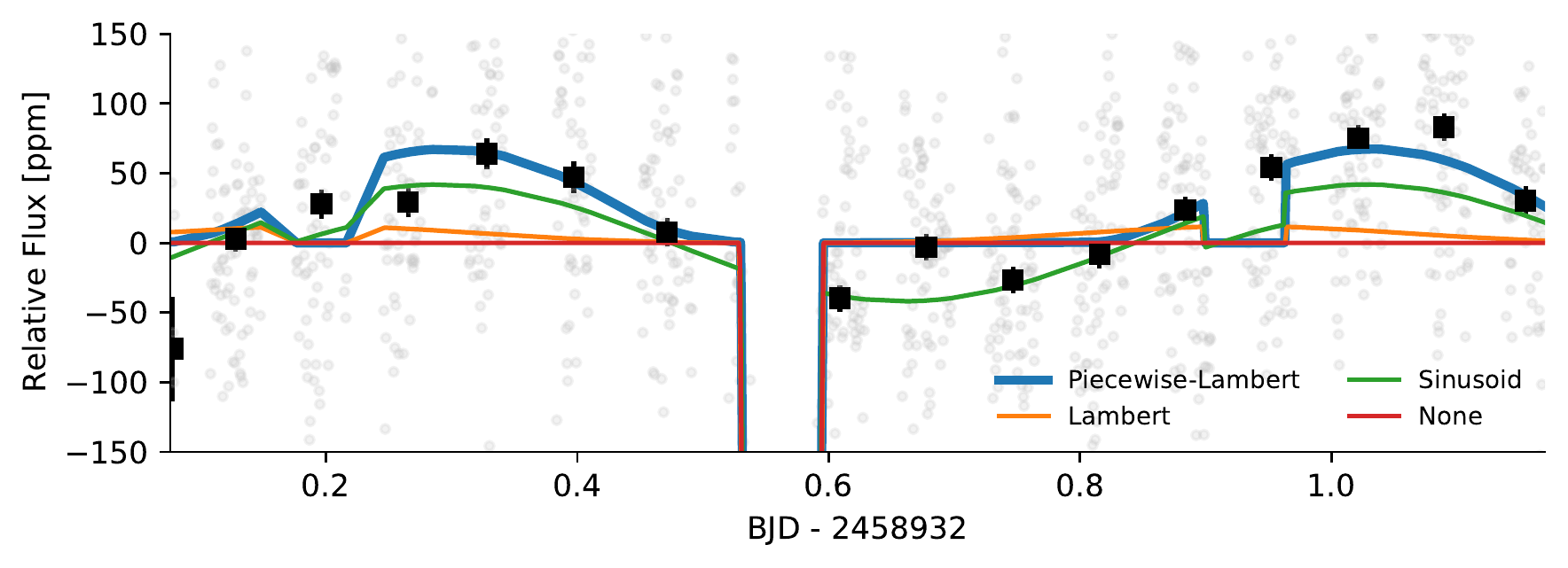}
    \caption{Comparison of the different models tested (colored curves), compared to the best-fit residuals obtained for the sinusoidal model (gray, and binned once per CHEOPS orbit in black).}
    \label{fig:my_label}
\end{figure*}

\section{Results} \label{sec:results}

Table~\ref{tab:transit} lists the best-fit transit and Piecewise-Lambertian model parameters. The ratio of radii $R_p/R_\star$ and impact parameter $b$ are consistent with the \spitzer observations by \citet{Demory2016a}.

The maximum-likelihood eclipse depth is ${19.8}^{+8.7}_{-9.0}$ ppm, or $2.2\sigma$ consistent with zero eclipse depth. This is consistent with the \citet{Kipping2020} claim of TESS occultation of depth $15.0 \pm 4.8$ ppm.

\begin{table}
\centering
\begin{tabular}{l c c}
Parameter & Piecewise-Lambert & Sinusoid \\ \hline
\multicolumn{3}{l}{\it Priors} \\
$R_\star~[R_\odot]$ & ${0.9415}^{+0.0099}_{-0.0100}$ & ${0.941}^{+0.010}_{-0.010}$ \\
$M_\star~[M_\odot]$ & ${0.906}^{+0.015}_{-0.015}$ & ${0.906}^{+0.015}_{-0.015}$ \\
$P$ [d] & $0.73654737^b$\\ \hline
\multicolumn{3}{l}{\it Fitted parameters}\\
$t_0$ [BJD$_{\rm TBD}$ - 2458932] & ${0.00042}^{+0.00031}_{-0.00032}$ & ${0.00031}^{+0.00032}_{-0.00033}$ \\
$u_1$ & ${0.132}^{+0.158}_{-0.093}$ & ${0.129}^{+0.157}_{-0.093}$ \\
$u_2$ & ${0.72}^{+0.15}_{-0.23}$ & ${0.73}^{+0.15}_{-0.21}$ \\
$R_p/R_\star$ & ${0.01693}^{+0.00035}_{-0.00035}$ & ${0.01696}^{+0.00033}_{-0.00034}$ \\
$b$ & ${0.323}^{+0.053}_{-0.068}$ & ${0.325}^{+0.051}_{-0.064}$ \\
Full amplitude [ppm] & ${71.3}^{+6.5}_{-6.5}$ & ${81.9}^{+7.9}_{-7.5}$ \\
$\xi_1$ & ${-0.66}^{+0.14}_{-0.12}$ & -- \\
$\xi_2$ & ${1.52}^{+0.15}_{-0.14}$ & -- \\
$\phi$ [rad] & -- & ${0.91}^{+0.10}_{-0.10}$ \\
\end{tabular}
\caption{Maximum-likelihood parameters for the transit model: mid-transit time $t_0$, ratio of planetary-to-stellar radius $R_p/R_\star$, impact parameter $b$, quadratic limb-darkening parameters $u_1, u_2$, full amplitude of the phase variations, and the western- and eastern-most longitudes of the region of the planet with lower reflectivity, $\xi_1$ and $\xi_2$. References: (a) \citet{vonBraun2011}; (b) \citet{Bourrier2018}. \label{tab:transit}}
\end{table}

\subsection{Comparison with PSF photometry} \label{sec:psf_comparison}

The analysis described above was conducted on the Data Reduction Pipeline (DRP) photometric products. To validate the results from DRP photometry, we also reduce and analyze PSF photometry constructed from the imagettes. The PSF photometric extraction is summarized in Appendix~\ref{sec:psf_appendix}.

We analyze the PSF photometry with the same procedure as outlined above for the DRP photometry. The differences in the two analyses include: (1) the detrending basis vectors for the PSF photometry include two principal components of the PSF shape changes in place of the ThermFront 2 vector (details in Appendix~\ref{sec:psf_appendix}); (2) the photometric cadence is shorter because photometry is extracted from the imagettes; (3) a smaller fraction of the total number of exposures is rejected due to the finer time sampling.

The same sequence of models is fit to the PSF photometry and we verify that the PSF photometry produces consistent (within 1-$\sigma$) measurements for each of the physical parameters in Table~\ref{tab:transit}. This validation step strengthens the claim that the flux variations are indeed astrophysical and their interpretation is robust against different photometric extraction techniques.

\section{Discussion} \label{sec:discussion}

Combining the optical CHEOPS observations with the existing optical and infrared observations creates a difficult puzzle. The CHEOPS optical light curves confirm that the flux variations may be variable in time, as suggested by the mismatch in phase and amplitude with the MOST observations of \citet[][see also Appendix~\ref{app:most}]{Sulis2019}. CHEOPS is scheduled to observe 55 Cnc e again, and the exquisite photometric precision demonstrated by CHEOPS on the single phase curve observation presented here suggests that we will verify the amplitude and timescales of the flux variations in the system. We have shown that CHEOPS can achieve sufficient precision to search for variability in the phase variations in the 55 Cnc system from one phase to the next, due to the exceptionally high-precision optical photometry from CHEOPS.

In the following subsections, we enumerate a series of mysteries left in the 55 Cnc system which future CHEOPS observations can interrogate.

\subsection{Planetary reflection}

In principle, the maximum-likelihood eclipse depth of $20 \pm 8$ ppm constrains the geometric albedo of the planet, which can be defined as:
\begin{equation}
    \delta_\mathrm{ecl} = A_g \left(\frac{R_p}{a}\right)^2
\end{equation}
if we neglect thermal emission in the CHEOPS bandpass. In practice however, the $1\sigma$ upper limit on the eclipse depth yields an uninformative upper limit on the geometric albedo $A_g < 0.98$. 

\subsection{Optical secondary eclipse}

The significant eclipse detected in the infrared and the lack of a clear eclipse in the optical presents another puzzle. The interpretation that the infrared eclipse is due to planetary light being occulted by the host star is challenged by the fact that the eclipse depth changed from one observation to the next, requiring the planet to change in size, albedo, thermal emission, or some combination of the three. Future observations will eventually measure any variability in the transit depth in the optical, which constrains the first scenario. Albedo variations could be detected by CHEOPS as variations in the optical eclipse depth, and will require further observation. The thermal emission could be further tested with large space-based infrared observatories.

The phase variation model with the maximum predictive power is the piecewise-Lambertian model of \citet{Hu2015} which describes an asymmetric albedo distribution on the planet. For self-consistency with this model, we have chosen to fix the eclipse depth to be equivalent to the full-amplitude of the phase variations at the mid-eclipse time. This could affect the Monte Carlo results in the following way: attempting to fit the significant phase variations with a high albedo will produce a strong secondary eclipse which is not observed, so the fit will negotiate a compromise between significant phase variations and an insignificant secondary eclipse.

\subsection{Properties of circumstellar or circumplanetary material}

Another hypothesis based on \citet{Demory2016a} states the planet is surrounded by material that is opaque in the optical but transparent in the infrared. Clumps of material could cause the variable flux modulation observed in the optical. This could occur if the circumstellar or circumplanetary dust has a narrow range of particle sizes.  Let the radius of the particle be $r$, the wavelength be $\lambda$ and $x=2\pi r/\lambda$. For CHEOPS, $x \gtrsim1$ implies $r \gtrsim 0.1$ $\mu$m. For 4.5 $\mu$m Spitzer, $x\lesssim 1$ implies $r \lesssim 0.7$ $\mu$m.  Collectively, the plausible range of particle radii is $0.1 \lesssim r \lesssim 0.7$ $\mu$m.  

\subsection{Star-planet interactions}

Interactions between the stellar and planetary magnetic fields could be responsible for excess flux emitted by the system, which could appear to vary on the orbital timescale of the planetary orbit \cite[see e.g.:][]{Jardine2008}. We can use order-of-magnitude estimates for the energy liberated by magnetospheric interactions in the star-planet system with the observed stellar magnetic field properties from \citet{Folsom2020} to estimate if sufficient energy is available to generate the observed phase variations with amplitude $\sim 80$ ppm. \citet{Zarka2007} gives a simple relation for the power $P_d$ dissipated by unipolar or dipolar satellite-magnetosphere interactions
\begin{equation}
    P_d \sim \epsilon \frac{vB^2}{\mu_0}\pi R_\mathrm{obs}^2 
\end{equation}
where $\epsilon$ is an unknown efficiency factor of order $0.1$, $v$ is the orbital velocity taken from \citet{Folsom2020}, $B$ is the local magnetic field at the planet e also from \citet{Folsom2020}, and $R_\mathrm{obs}$ is the obstacle radius which we take as the magnetospheric radius $R_m$ defined in \citet{Jardine2008}. We express the power dissipated $P_d / L$ as a ratio with the luminosity of the star $L = 0.582 L_\odot$ \citep{vonBraun2011}, finding $P_d / L \approx 10^{-10}$ for a planetary magnetic field strength of order $1$ G. Given that the full-amplitude of the CHEOPS phase variations $\sim 80$ ppm is five orders of magnitude larger than the energy budget for magnetospheric interactions, unipolar or dipolar interactions likely are not solely responsible for the phase variations in the system.

To further test the star-planet interaction hypothesis, time-resolved simultaneous observations of the phase curve and spectroscopic activity indicators may be necessary. If the star has chromospheric hot spots induced by the planet as imagined by \citet{Folsom2020}, those spots may emit excess flux in the CaII H \& K lines, which could be observed with ground-based spectroscopy. If the chromospheric activity indeed varies with the orbital phase of the planet, and also varies on longer timescales from one planetary orbit to another, the star-planet interaction model might succinctly explain the flux variations, though no such evidence for line variations was found by \citet{Ridden-Harper2016}.

\section{Conclusion} \label{sec:conclusion}

CHEOPS observations confirm that the 55 Cnc system varies in flux over the orbital period of planet e. The origin of these flux variations is unclear. Two scenarios ruled unlikely by this work are that the flux variations are the result of reflection from the planet's surface (observed amplitude is too large and asymmetric), or that magnetospheric interactions are inducing excess emission from the star (observed amplitude is too large). A surviving hypothesis for explaining the optical and infrared observations is that dust is orbiting either the star or planet, obscuring the secondary eclipse in the optical but not the infrared.

Though the observations of 55 Cnc e may be enigmatic, it is clear that CHEOPS is performing as predicted \citep{Hoyer2020,Futyan2020,Deline2020}. We have shown that CHEOPS is a photometrically stable observatory capable of measuring the $71.5 \pm 6.5$ ppm variability of the 55 Cnc system in a single visit. CHEOPS is currently collecting more high-precision phase curves for 55 Cnc and several other bright exoplanet systems, which will be the subjects of future publications.

\begin{acknowledgements}
    This manuscript benefited from the feedback provided by an anonymous referee. We gratefully acknowledge the open source software which made this work possible: \texttt{astropy} \citep{Astropy2013, Astropy2018}, \texttt{ipython} \citep{ipython}, \texttt{numpy} \citep{numpy}, \texttt{scipy} \citep{scipy}, \texttt{matplotlib} \citep{matplotlib}, \texttt{emcee} \citep{Foreman-Mackey2013}, \texttt{batman} \citep{Kreidberg2015}, \texttt{PyMC3} \citep{pymc3}, \texttt{arviz} \citep{arviz_2019}.
    
    This research made use of \texttt{exoplanet} and its dependencies \citep{exoplanet:agol20,exoplanet:astropy13,exoplanet:astropy18,exoplanet:exoplanet, exoplanet:kipping13, exoplanet:luger18, exoplanet:pymc3, exoplanet:theano}.

    This work has been carried out in the framework of the PlanetS National Centre of Competence in Research (NCCR) supported by the Swiss National Science Foundation (SNSF). 

    CHEOPS is an ESA mission in  partnership with Switzerland with important contributions to the payload and the ground segment from Austria, Belgium, France, Germany, Hungary, Italy, Portugal, Spain, Sweden, and the United Kingdom. The Swiss participation to CHEOPS has been supported by the Swiss Space Office (SSO) in the framework of the Prodex programme and the Activit\'es Nationales Compl\'ementaires (ANC), the Universities of Bern and Geneva as well as of the NCCR PlanetS and the Swiss National Science Foundation.

    This work benefited from support from the Swiss National Science Foundation (PP00P2-163967 and PP00P2-190080).

    This project has received funding from the European Research Council (ERC) under the European Union’s Horizon 2020 research and innovation programme (project {\sc Four Aces}; grant agreement No~724427).
    
    This work was supported by FCT - Funda\c{c}\~ao para a Ci\^encia e a Tecnologia through national funds and by FEDER through COMPETE2020 - Programa Operacional Competitividade e Internacionaliza\c{c}\~ao by these grants: UID/FIS/04434/2019; UIDB/04434/2020; UIDP/04434/2020; PTDC/FIS-AST/32113/2017 \& POCI-01-0145-FEDER-032113; PTDC/FIS-AST/28953/2017 \& POCI-01-0145-FEDER-028953; PTDC/FIS-AST/28987/2017 \& POCI-01-0145-FEDER-028987. S.C.C.B. and S.G.S. acknowledge support from FCT through FCT contracts nr. IF/01312/2014/CP1215/CT0004, IF/00028/2014/CP1215/CT0002. O.D.S.D. is supported in the form of work contract (DL 57/2016/CP1364/CT0004) funded by national funds through Funda\c{c}\~ao para a Ci\^encia e Tecnologia (FCT). XB, SC, DG, MF and JL acknowledge their roles as ESA-appointed CHEOPS science team members.
    
    ACC and TW acknowledge support from STFC consolidated grant number ST/R000824/1.
    
    This project was supported by the CNES. SH gratefully acknowledges CNES funding through the grant 837319.
    
    PM acknowledges support from STFC consolidated grant number ST/M001040/1.

    KGI is the ESA CHEOPS Project Scientist and is responsible for the ESA CHEOPS Guest Observers Programme. She does not participate in, or contribute to, the definition of the Guaranteed Time Programme of the CHEOPS mission through which observations described in this paper have been taken, nor to any aspect of target selection for the programme.

\end{acknowledgements}

\appendix

\section{Posterior distributions} \label{app:corner}

\begin{figure*}
    \centering
    \includegraphics[width=\textwidth]{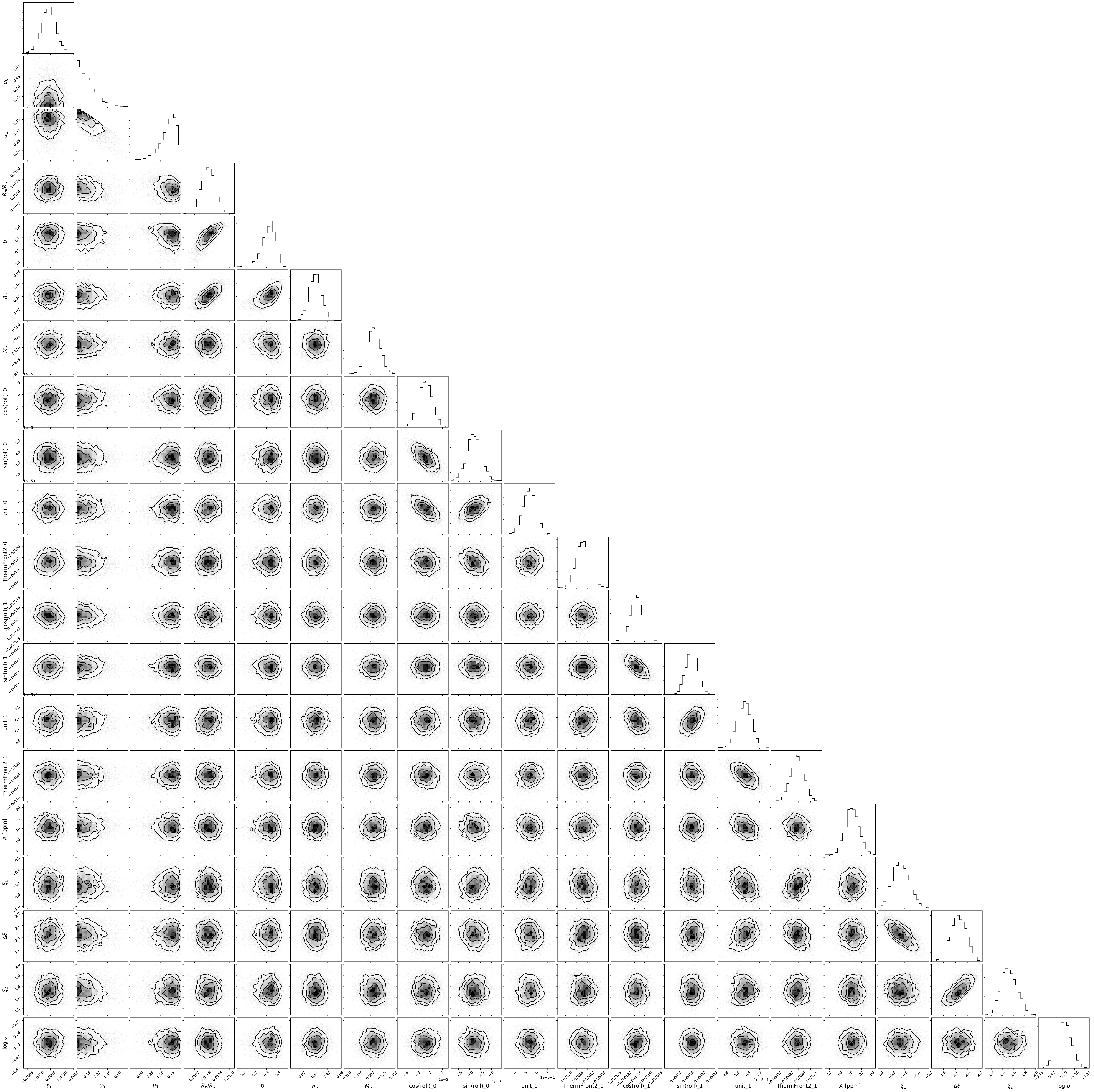}
    \caption{Posterior distributions and joint correlations between all free parameters in the joint fit to the In-Orbit Commissioning transit and phase curve observations sampled with the No U-Turn Sampler. The parameters include: the mid-transit time $t_0$, limb-darkening parameters $u_0, u_1$, the ratio of planetary to stellar radius $R_p/R_\star$, the impact parameter $b$, the eclipse depth $\delta_\mathrm{ecl}$; two sets of detrending vector weights for the following vectors, one per visit: the ThermFront 2 temperature sensor reading, the cosine and sine of the roll angle, and a unit vector; the amplitude of the piecewise-Lambertian variations $A$; and the start and stop longitudes of the lower-reflectivity region $\xi_1$ and $\xi_2$. The last parameter, $\log \sigma$ is the natural logarithm of the flux uncertainty for each measurement. }
    \label{fig:corner}
\end{figure*}

In the corner plot in Figure~\ref{fig:corner} there are two sets of these basis vectors included in the fit, one per CHEOPS visit, corresponding to the In-Orbit Commissioning (IOC) transit observation and the phase curve observation, respectively.

\section{PSF Photometry} \label{sec:psf_appendix}

\begin{figure}
    \centering
    \includegraphics[width=0.4\textwidth]{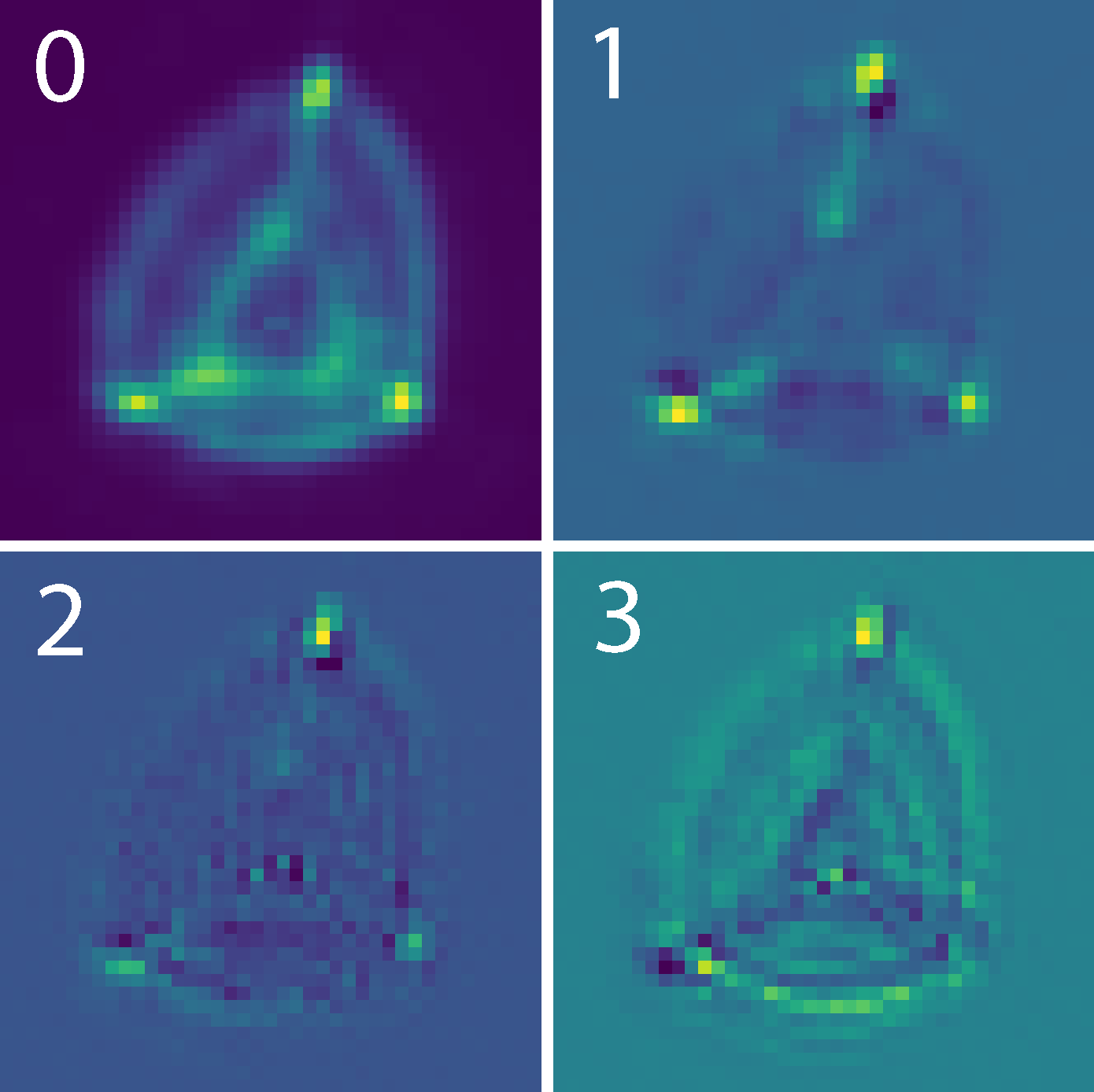}
    \caption{The four principal components with the highest eigenvalues (from the PCA of the PSFs).}
    \label{fig:top_four_psf_pcs}
\end{figure}

We used both imagettes and subarrays of 55\,Cnc to measure how the PSF changes with time by deriving independent PSFs for each of the visit’s 16 CHEOPS orbits. For each exposure, the pixel values were normalised and listed with coordinates relative to the center of the PSF. Each CHEOPS orbit produced about 75 subarray images and 750 imagettes, with centres dithered by a fraction of a pixel. The table of pixel values and their relative pixel coordinates was then fit with a 2D-spline with knots located in a grid of the same resolution as the pixel grid. Sigma-clipping was used to remove deviating pixels (from e.g.\ cosmic ray hits). To find what parts of the PSF changes with time, a principal component analysis (PCA) was performed on the spline coefficients of the 16 derived PSFs. The 4 most important components are illustrated in Fig.~\ref{fig:top_four_psf_pcs}.

\begin{figure}
    \centering
    \includegraphics[width=0.4\textwidth]{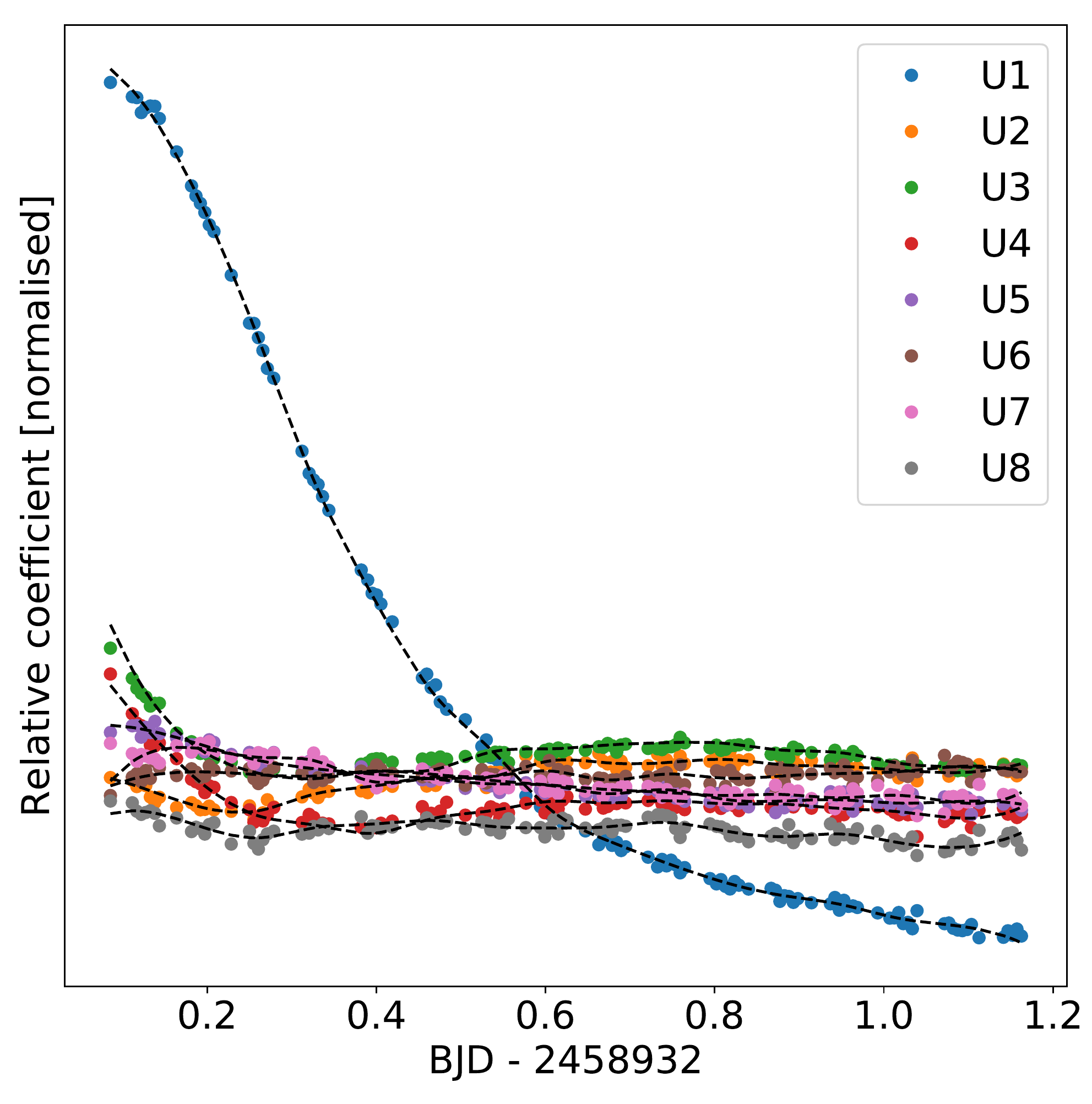}
    \caption{The relative coefficients of the U1 to U8 components from the PCA of the PSFs as a function of time. The roll angle dependence has been removed in this plot.}
    \label{fig:top_eight_pcs_time}
\end{figure}

\begin{figure}
    \centering
    \includegraphics[width=0.4\textwidth]{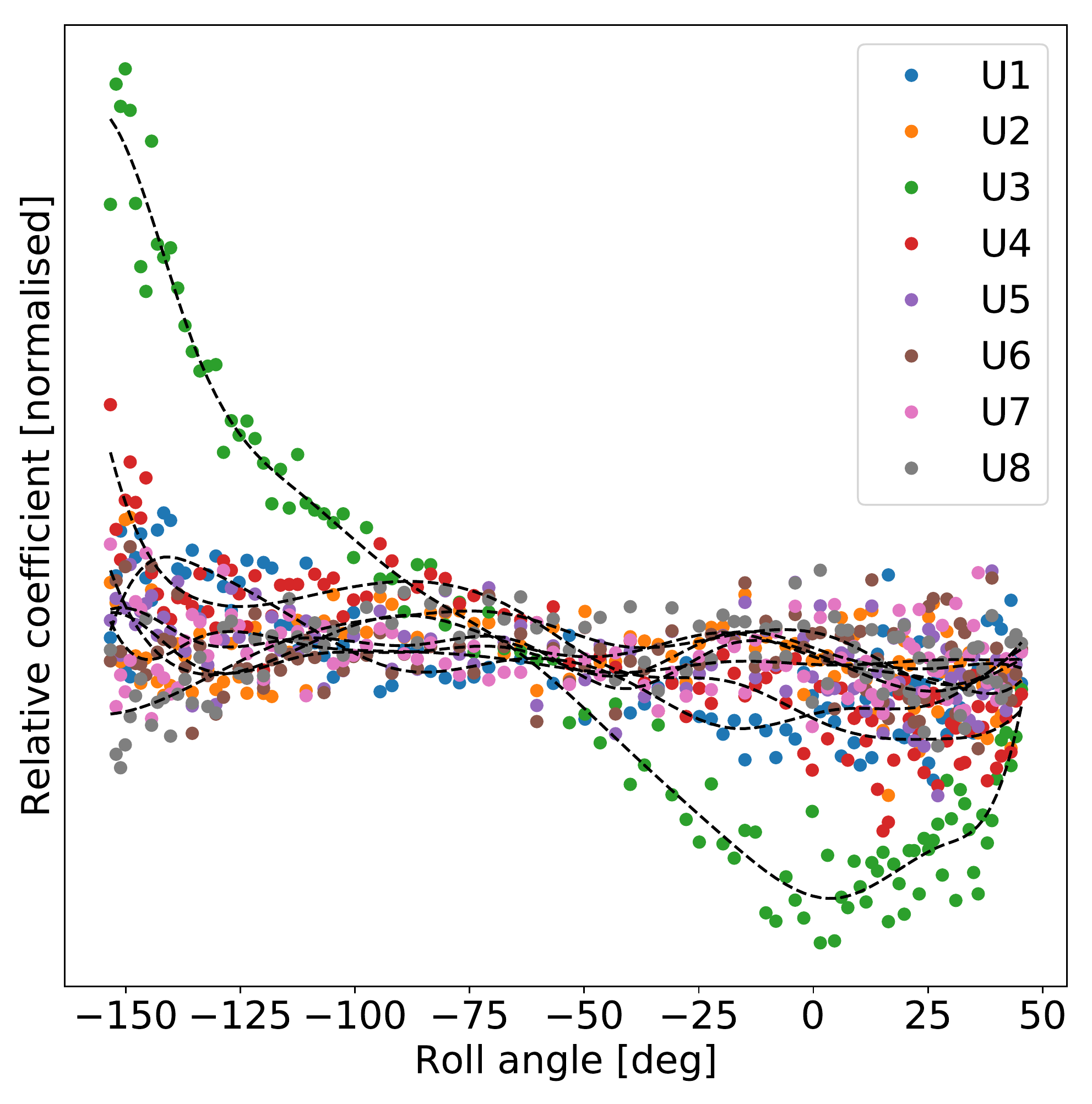}
    \caption{The relative coefficients of the U1 to U8 components from the PCA of the PSFs as a function of roll angle. The long-term time dependence from Fig.~\ref{fig:top_four_psf_pcs} has been removed in this plot, to reduce the scatter and better see the roll angle trend.}
    \label{fig:top_eight_pcs_roll}
\end{figure}

For each imagette, a best-fit PSF was produced by combining the first 10 principal components (named U0 to U9, where U0 is the average PSF). The relative importance of each component can then be followed in time to track PSF changes. Fig.~\ref{fig:top_eight_pcs_time} shows how U1 to U8 vary in time, where U1 is clearly dominating the change with time. In this figure any roll angle dependent variation has been removed and the measurement points have been binned by a factor 100. In Fig.~\ref{fig:top_eight_pcs_roll} we see the component coefficients as a function of roll angle, where the long-term dependence from Fig.~\ref{fig:top_eight_pcs_time} has been removed and the measurement points again binned by a factor of 100, but this time in roll angle. The roll angle dependence is generally weak, except for U3. The photometric precision improves only slightly by fitting an increasingly accurate PSF. Going from one principal component to 3 only improves the photometric precision by 2.5\% as measured by the point-to-point mean absolute deviation. The precision keeps improving slowly another 0.5\% until 10 components are used, and decreases slowly after that, likely due to the over-fitting of noise. The main advantage of decomposing the PSF into principal components is thus not necessarily to improve the noise, but to follow how the PSF changes over time and enable correction for long-term trends (in particular the ramp; see Fig.~\ref{fig:raw}).

\begin{figure}
    \centering
    \includegraphics[width=0.45\textwidth]{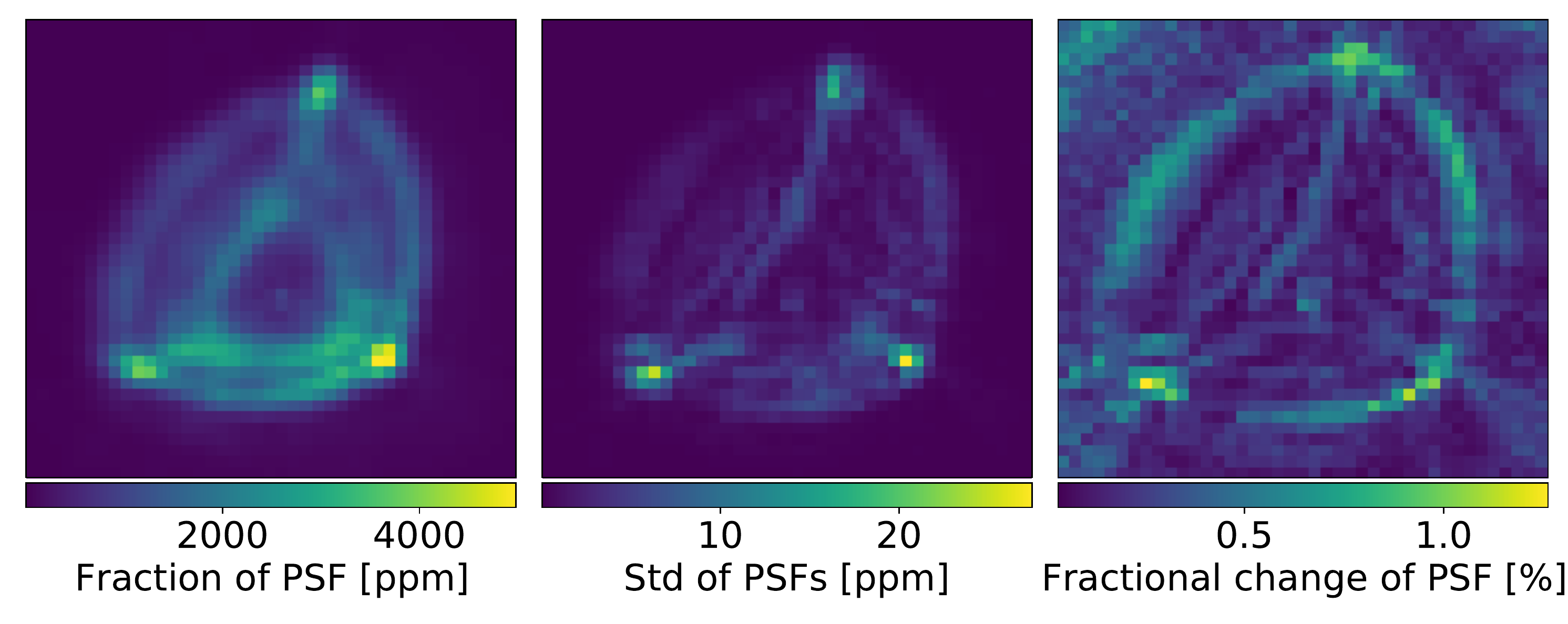}
    \caption{The CHEOPS PSF and its changes over the 55\,Cnc\,e visit. The left panel shows the mean PSF, the central panel shows the standard deviation of the PSF per pixel over the duration of the visit, and the right panel shows the fractional change of the PSF per pixel.}
    \label{fig:psf_changes}
\end{figure}

Fig.~\ref{fig:psf_changes} shows the mean PSF, the standard deviation of pixels in the PSF over the visit, and the relative change of the PSF. The three bright corner spots are dominating the change with variations of up to 6\%, while the other parts of the PSF are generally stable on a fraction of a percent level.

\begin{table}
\centering
\begin{tabular}{lcr}
Parameter & DRP & PSF \\ \hline
$t_0$ [BJD$_{\rm TBD}$ - 2458932] & ${0.56284}^{+0.00028}_{-0.00028}$ & ${0.56258}^{+0.00025}_{-0.00025}$ \\
$R_p/R_\star$ & ${0.01719}^{+0.00028}_{-0.00027}$ & ${0.01718}^{+0.00026}_{-0.00026}$ \\
$b$ & ${0.360}^{+0.036}_{-0.044}$ & ${0.311}^{+0.041}_{-0.050}$ \\
Full amplitude [ppm] & ${70.9}^{+6.7}_{-6.4}$ & ${70.4}^{+4.6}_{-4.4}$ \\
$\xi_1$ [rad] & ${-0.66}^{+0.14}_{-0.13}$ & ${-0.58}^{+0.10}_{-0.11}$ \\
$\xi_2$ [rad] & ${1.52}^{+0.16}_{-0.14}$ & ${1.549}^{+0.086}_{-0.088}$ \\
\end{tabular}
\caption{Comparison of the best-fit transit model parameters for the DRP and PSF photometric reductions. Limb-darkening parameters were fixed across both reductions to $u_1, u_2 = 0, 0.75$. \label{tab:drp_psf_comparison}}
\end{table}

\begin{figure*}
    \centering
    \includegraphics{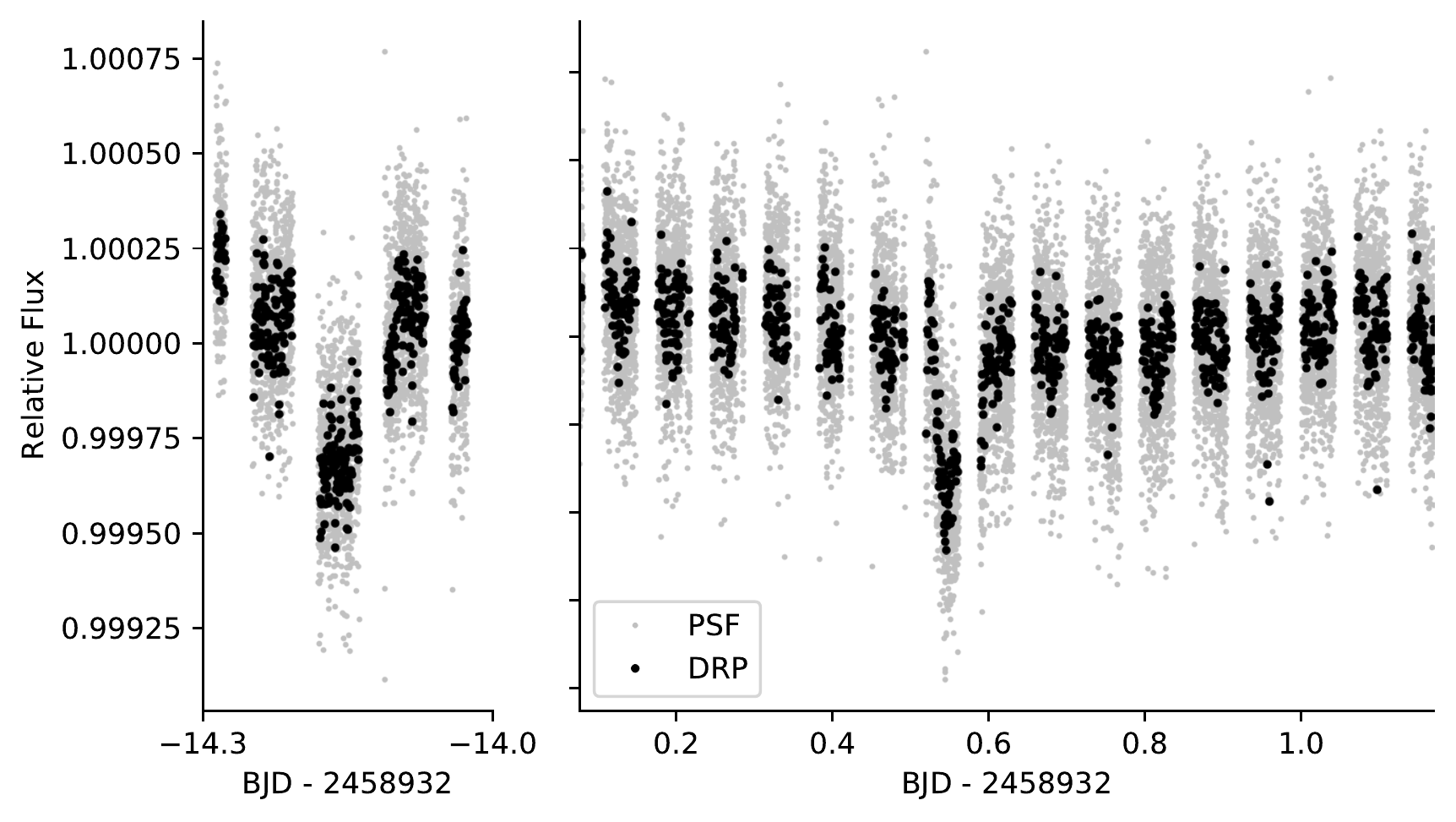}
    \caption{Comparison of the CHEOPS Data Reduction Pipeline (DRP) aperture photometry (black) with the PIPE PSF photometry (gray). The PSF photometry is computed on the imagettes and therefore has a higher cadence than the DRP photometry. The In-Orbit Commissioning (IOC) transit observations are on the left, and the first 55 Cnc e phase curve observation is on the right.}
    \label{fig:psf_drp}
\end{figure*}

\section{Comparison with MOST photometry} \label{app:most}

\begin{figure}
    \centering
    \includegraphics[scale=0.85]{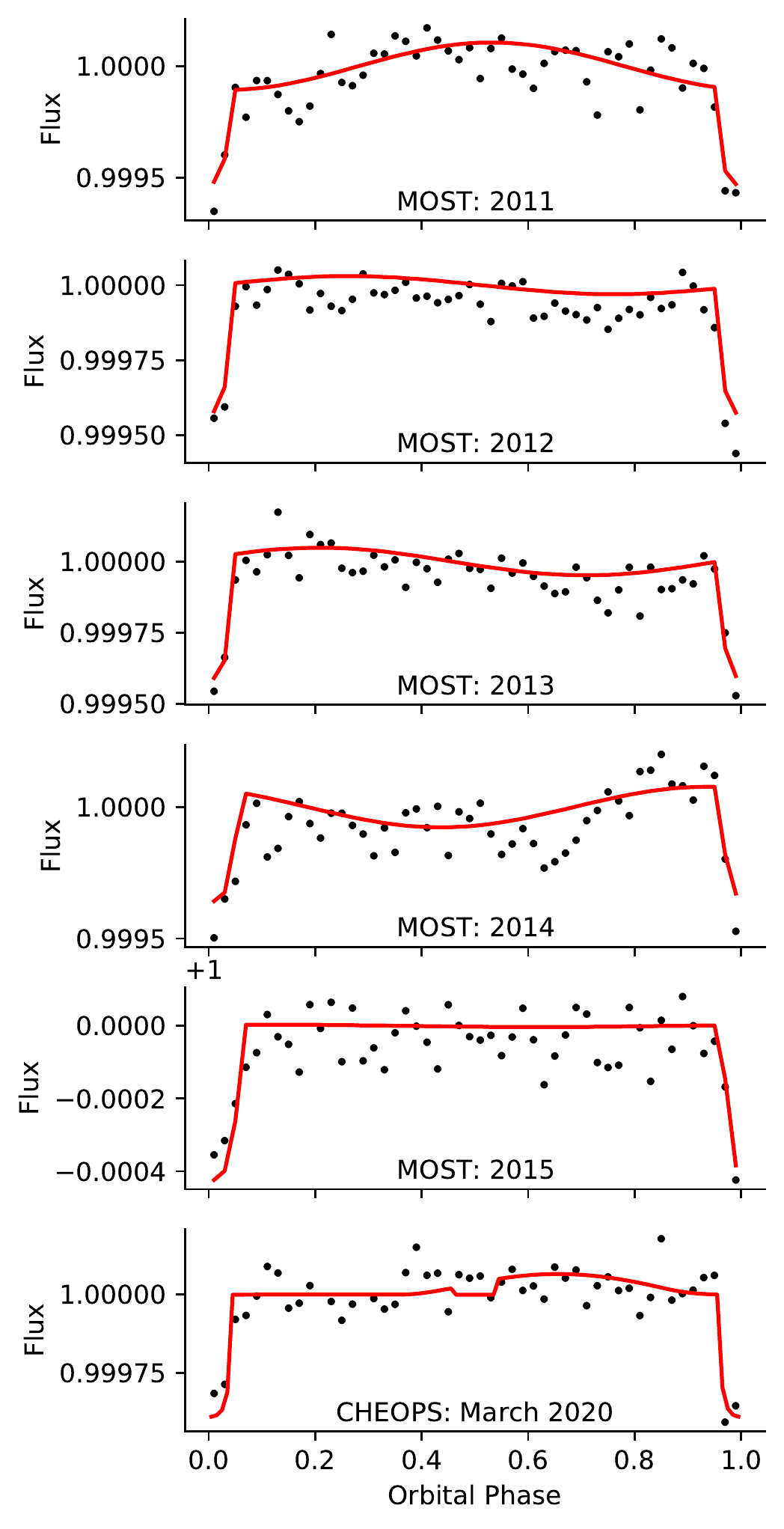}
    \caption{ Comparison of the yearly MOST photometry from \citet{Sulis2019} with the CHEOPS observations in the bottom panel. The red curves for each MOST observation is a simple sinusoid and transit model, while the red curve for the CHEOPS observations represents the piecewise-Lambertian model.}
    \label{fig:MOST}
\end{figure}

We plot the CHEOPS photometry of 55 Cnc alongside the yearly binned photometry from \citet{Sulis2019} in Figure~\ref{fig:MOST}. The apparent variations in the phase curve shape and amplitude observed with MOST seem to continue into 2020 from the CHEOPS observations.

\bibliographystyle{aa} 
\bibliography{bibliography} 
\end{document}